\documentclass[pre,showpacs,twocolumn,preprintnumbers,amsmath,amssymb]{revtex4}

\usepackage{graphicx}
\usepackage{dcolumn}
\usepackage{bm}
\usepackage{parskip}
\usepackage{subfigure}
\usepackage{amssymb}
\usepackage{amsmath}
\usepackage{amssymb}
\usepackage{graphicx}

\def\be{\begin{eqnarray}}
\def\ee{\end{eqnarray}}
\def\be{\begin{equation}}
\def\ee{\end{equation}}
\usepackage{color}

\begin{document}
\title{Pronounced minimum of the thermodynamic Casimir forces  \\
of O(${\bf n}$) symmetric  film systems:  Analytic theory
}

\author{Volker Dohm}

\affiliation{Institute for Theoretical Physics, RWTH Aachen
University, D-52056 Aachen, Germany}

\begin{abstract}
Thermodynamic Casimir  forces of film systems in the O$(n)$ universality classes with Dirichlet boundary conditions are studied below bulk criticality. Substantial progress is achieved in resolving the long-standing problem of describing analytically the pronounced minimum of the scaling function observed experimentally in $^4$He films $(n=2)$ by Garcia and  Chan [Phys. Rev. Lett. ${\bf 83}, 1187 \;(1999)$] and in Monte Carlo simulations for the three-dimensional Ising model ($n=1$) by O. Vasilyev et al. [Europhys. Lett. ${\bf 80}, 60009 \;(2007)$]. Our finite-size renormalization-group approach describes the film systems as the limit of finite-slab systems with vanishing aspect ratio. This yields excellent agreement with the depth and the position of the minimum for $n=1$ and semiquantitative agreement with the minimum for $n=2$. Our theory also predicts a pronounced minimum for the $n=3$ Heisenberg universality class.
\end{abstract}

\maketitle

Thermodynamic Casimir forces occur in a large variety of confined condensed-matter systems \cite{kardar} and have attracted the interest of many theoretical and experimental researchers over past decades, including very recently \cite{fisher78,night,KrDi92,garcia,zandi2004,wil-1,hucht,zandi2007,vasilyev2007,maciolek,dohm2009,hasenbusch2010,kastening-dohm,biswas2010,diehl2012,dohm2013,abraham}. Of particular interest are O$(n)$ symmetric film systems where both long-range Goldstone and critical fluctuations are the physical origin of such Casimir forces. One of the most prominent systems is superfluid $^4$He $(n=2)$ where the Casimir force causes a surprising and as yet unexplained effect close to the superfluid transition: a pronounced minimum of the Casimir force scaling function at a temperature $T_{\text min}$ as observed experimentally by a thinning of liquid $^4$He films \cite{garcia}. This effect has been confirmed by Monte Carlo (MC) simulations for the $XY$ model $(n=2)$ \cite{hucht,vasilyev2007,hasenbusch2010}, and similar minima were found in  MC data for the Ising model $(n=1)$  \cite{vasilyev2007} and in a numerical analysis of the O$(n)$ $\varphi^4$ model with free boundary conditions (BC) in the large - $n$ limit \cite{diehl2012}.  In all cases, $T_{\text min}$ is found to be below bulk  $T_c$ and, for $n=1,2$, above the film critical temperature $T_{c,\text{film}}$. This calls for an explanation of the minima that is not specific to the superfluid transition and the $XY$ universality class and is largely unrelated to the existence of and the crossover to a Goldstone regime at low temperatures.

An early  renormalization-group (RG) description of the thermodynamic Casimir effect in $d=4-\varepsilon$ dimensions \cite{KrDi92} covers essentially only the region above bulk criticality where this effect is quite small and where no indication of the large minimum below bulk criticality of $^4$He is recognizable. Subsequent theoretical work is based on mean field (MF) theory \cite{zandi2007,biswas2010,maciolek} which, however, is not capable of making a prediction of the depth of the minimum because of the strong dependence on an undetermined nonuniversal parameter. A RG improved version of MF theory  \cite{zandi2007} yields a minimum that is roughly five times deeper than the experimentally measured minimum.
Recently an analytic RG calculation of the minimum  of $O(n)$ symmetric systems  in a $L_\parallel^2 \times L$ slab geometry with periodic BC and finite aspect ratio $\rho=L/L_\parallel$  \cite{dohm2009,dohm2013} was found to be in agreement with MC data \cite{vasilyev2007,hucht2011,dan-krech} but the position and the depth of the minima are rather far from those of the minimum in real $^4$He films \cite{garcia} which requires a description with Dirichlet BC because of the vanishing of the order parameter at the boundaries.

In this paper we develop an analytic theory of the Casimir force that
is in substantially improved agreement with the observed minima of systems with free or Dirichlet BC. Our approach is focused at the outset on the important  region $T_{c,\text{film}} < T < T_c$.
It is based on the physical fact that the {\it disordered} phase of the system for $n=1$ and $n=2$ includes the entire region above $T_{c,\text{film}}$ rather than only the region above bulk $T_c$ which enables us to develop a finite-size theory for general $n$  {\it below} $T_c$ without encountering problems due to Goldstone modes.
We also predict a minimum for the $(n=3)$ Heisenberg universality class. Our theory with Dirichlet BC should also be an appropriate basis for describing Casimir forces in superconducting films \cite{wil-1} and finite-size effects in confined magnetic materials provided that the theory includes the effects of lattice anisotropy \cite{dohm2008}.

We start from the O$(n)$ symmetric $\varphi^4$ Hamiltonian
\begin{eqnarray}
\label{Hamiltonian}H &=& \int_{V} d^d x
\big[\frac{r_0} {2} \varphi^2 +  \frac{1} {2} (\nabla
\varphi)^2 +
 u_0 (\varphi^2)^2   \big]
\end{eqnarray}
where $\varphi({\bf x})$ is an $n$-component field in a $d$ dimensional $L_\parallel^{d-1} \times L$ slab geometry with a  finite volume $V=L_\parallel^{d-1}L $. We consider periodic BC in the $d-1$ "horizontal" directions but Dirichlet BC in the $d^{th}$ "vertical" direction. Accordingly $\varphi({\bf x})\equiv \varphi({\bf y},z)$ is represented as $\varphi({\bf x})=\sqrt 2\sum_{{\bf n},m}\hat \varphi_{{\bf n},m} e^{i{\bf p}{\bf y}}\sin(q z)$ where the sum $\sum_{{\bf n},m}$ runs over $(d-1)$ - dimensional ${\bf p}$ vectors with components $p_\alpha=2\pi n_\alpha/L_\parallel$, $\alpha= 1,2, ..., d-1$, with integers $n_\alpha=0, \pm 1,\pm 2,...,$ and over wave numbers $q=\pi m/L$
with integers $m= 1,2,...$ (up to some cutoff $\Lambda$). Our system differs from the periodic slabs studied previously \cite{dohm2013} in that now there exist surface contributions to the free energy  and   an {\it inhomogeneous} lowest  mode $\psi(z)=\Phi \sqrt 2 \sin(\pi z/L)$ with $\Phi \equiv \hat\varphi_{{\bf 0},1}$  which implies an enhanced four-point coupling of the lowest-mode Hamiltonian and an (unrenormalized) shift $\propto L^{-2}$ of the film transition. As shown below, our approach succeeds in renormalizing this shift for Dirichlet BC and in predicting a finite temperature range $T_{c,\text{film}} < T < T_c$, in good agreement with experiment and MC data, whereas this temperature range is not captured in the film limit   $\rho \to 0$ of the RG theory of Ref. \cite{dohm2013} for periodic BC where all three temperatures $T_{c,\text{film}},  T_{\text{min}}$, and $T_c$ coincide.

Our applications will be focused on  Ising-like and XY-like systems.
For $\rho \to 0$, they undergo a phase transition for $ n=1,d>2$ and for $n=2,d\geq 3$  at a finite temperature $0<T_{c,\text{film}}(L) < T_c$. For $n=2, d=3$, this is a Kosterlitz-Thouless transition.  No finite $T_{c,\text{film}}$ exists for $n>2$ in $d\leq3$ dimensions.
The Casimir force per unit area $F_{{\text Cas}}=-\partial [Lf^{{\text ex}}]/\partial L$
can be derived from the excess free energy density (divided by $k_BT$) $f^{{\text ex}}=f-f_b$ where $f(T,L,L_\parallel)= -V^{-1}\ln \int {\cal D} \varphi \exp(-H)$ and  $f_b \equiv \lim_{V\rightarrow \infty}f$ are the free energy densities of the finite and bulk system, respectively. One expects that, for isotropic systems near $T_c$ and for large $L$ and $L_\parallel$,  $F_{{\text Cas}}$ can be written in a scaling form \cite{pri}
\begin{eqnarray}
\label{scaling}
F_{{\text Cas}}(t, L, L_\parallel)=L^{-d} X(\tilde x,\rho)
\end{eqnarray}
with the scaling variable $ \tilde x=t(L/\xi_{0+})^{1/\nu}$, $t=(T-T_c)/T_c$ where $\xi_{0+}$ is the  amplitude of the bulk correlation length   {\it above} $T_c$.  We derive the scaling function $X(\tilde x,0)$ above $T_{c,\text{film}}$ for general $n$ without any adjustment of parameters.

Unlike earlier theories \cite{KrDi92,wil-1,biswas2010} for film ($\rho =0$) geometry, our strategy is to describe the film system as the limit $\rho \to 0$ of the finite-slab ($\rho >0$) systems. This differs from \cite{dohm2009,dohm2013} whose applicability near $T_c$ is restricted to $\rho >0$. We first present our approach for $n=1$. We decompose $\varphi({\bf x}) =  \psi(z)+\hat \varphi({\bf x})$ with the higher-mode fluctuations $\hat \varphi({\bf x}) = \sum'_{{\bf n},m}\sqrt 2\hat \varphi_{{\bf n},m} e^{i{\bf p}{\bf y}}\sin(q z)$ where the sum $\sum'_{{\bf n},m}$ does not include the lowest mode $({\bf 0},1)$.  Accordingly we decompose  $ H=H_0+H^{(2)}+H^{(3)}+H^{(4)} $,
\begin{eqnarray}
\label{lowHamilton}
H_0 (\Phi^2) = V
\left[\frac{1}{2} (r_0 +\pi^2/L^2)\Phi^2 + \frac{3}{2}u_0 \Phi^4  \right],
\end{eqnarray}
\begin{eqnarray}
\label{higher Hamilton}
&&H^{(2)} (\Phi, \hat \varphi)={\sum_{{\bf n},m}}'\Big\{ \frac{1}{2}[ \tilde r(\Phi^2)+ \mathbf p^2 + q^2]\hat\varphi_{{\bf n},m}\hat\varphi_{{\bf -n},m} \nonumber \\&&+ b(\Phi^2)[ \hat\varphi_{{\bf n},m}\hat\varphi_{{\bf -n},m}\delta_{m,1}-\hat\varphi_{{\bf n},m}\hat\varphi_{{\bf -n},m+2}\nonumber \\&&-\hat\varphi_{{\bf n},m}\hat\varphi_{{\bf -n},m-2}]\Big\} -w(\Phi)\hat\varphi_{{\bf 0},3},
\end{eqnarray}
with  $b(\Phi^2)=3u_0\Phi^2$, $w(\Phi)=2u_0\Phi^3$, and the "longitudinal" parameter $ \tilde r(\Phi^2) = r_0+12u_0 \Phi^2$. The interesting aspect here is the treatment of {\it finite-size} effects on the basis of $H^{(2)}$ and  $H^{(3)}$ $\sim$ $O(u_0\Phi\hat\varphi^3)$ {\it below} $T_c$ whereas only {\it surface} properties {\it above} $T_c$ were treated previously \cite{dohm1989} on the basis of $H^{(2)}$ and $H^{(4)}$ $\sim$ $O(u_0\hat\varphi^4)$.
After integration over $\hat\varphi$, we obtain the  free energy density
\begin{eqnarray}
\label{bare-free}
f = f_0 - \frac{1}{V}\ln \Big\{ \int^{\infty}_{-\infty}  d \Phi \exp \left[-H_0(\Phi^2)- \Gamma(\Phi^2)\right]\Big\}  ,
\end{eqnarray}
\begin{eqnarray}
\label{Gamma}
\Gamma(\Phi^2) &&= {\sum_{{\bf n},m}}'\Big\{\frac{1}{2}\ln a_{{\bf n},m}(\tilde r, b) - \frac{ 2 b^2}{a_{{\bf n},m}(\tilde r, b)\;a_{{\bf n},m+2}(\tilde r, b)}\Big\}
\nonumber\\&&-6u_0\Phi w \frac{1}{a_{{\bf 0},3}}\Big\{{\sum_{\bf n}}'\frac{1}{a_{{\bf n},1}(\tilde r, b)}-{\sum_{\bf n}}\frac{1}{a_{{\bf n},2}(\tilde r, b)}\Big\}
\end{eqnarray}
apart from contributions of $O(b^3,w^2,u_0)$, with $a_{{\bf n},m}(\tilde r, b)=\tilde r + 2b\delta_{m,1}+4\pi^2{\bf n}^2/L_\parallel^2+\pi^2m^2/L^2$. The sum $\sum'_{{\bf n}}$ does not include ${\bf n}={\bf 0}$. The constant $f_0$ is independent of $r_0$ and $u_0$.
The main contribution of the integration over $ \Phi$ comes from the region around $\Phi^2\approx M_0^2$ where $M_0^2=\int^{\infty}_{-\infty} d  \Phi  \Phi^2\exp \left[-H_0(\Phi^2)\right]/\int^{\infty}_{-\infty}  d \Phi \exp \left[-H_0(\Phi^2)\right]$ is the lowest-mode average. This provides the justification for approximating $\Gamma(\Phi^2)$ by $\Gamma(M_0^2)$. The structure of our $\Gamma(M_0^2)$ is considerably more complicated than that of the sum ${\sum_{\bf k\neq0}} \ln (\tilde r + \mathbf k^2)$ of Ref. \cite{dohm2013} for periodic BC. In particular, our bulk limit of $f$  below $T_c$ differs from that for periodic BC \cite{dohm2013} because of the enhancement factor 3/2 in $H_0$ and the term  $\propto b(M_0^2)^2$ in $\Gamma(M_0^2)$.

Since the smallest value of $q$ is ${\it finite}$, namely $\pi/L $,  $f$ is an analytic function of $r_0$  at  $r_0=0$  for $\rho\geq0$. On the level of the unrenormalized theory, the analyticity for $\rho=0$  extends down to the film transition temperature at $r_0 = - \pi^2/L^2$. The focus of our theory is the (renormalized counterpart of the) range $r_0> -\pi^2/L^2$. As shown below, this range fully includes the minimum of the Casimir force scaling function where this function is nonsingular, in contrast to the MF results \cite{zandi2007,biswas2010,maciolek}.

According to our concept of analyzing the range above the film transition, we have performed an exact analytic calculation of $\Gamma(M_0^2)$ not only for $\tilde r(M^2_0) \equiv \hat r > 0$ but also {\it for its full range of existence} $\hat r > -\pi^2/L^2$ for finite  $L\gg\Lambda^{-1}, L_{\parallel}\gg\Lambda^{-1}$ and arbitrary $\rho>0$ above and below $T_c$ in $2<d<4$ dimensions including the limits $L \to \infty, L_\parallel \to \infty$.  This differs from earlier calculations \cite{KrDi92} of two-loop sums  that were restricted to $r_0>0$ and $\rho=0$. We apply this calculation to the excess free energy  above $T_{c,\text{film}}(L)$ for $0< \rho \ll 1$. The result above $(+)$ and below $(-)$ bulk $T_c$ reads $f^{{\text ex}}(r_0,L,L_\parallel)=f_s(\hat r,L,L_\parallel)- f^\pm_{b,s}(r_0)$ with the unrenormalized singular parts
\begin{eqnarray}
\label{excess-free-energy} f_s(\hat r,L,L_\parallel)&&= - \frac{ A_d}{ d\bar r^{\varepsilon/2}} \Bigg\{\frac{ \hat r^2}{\varepsilon}  +  \frac{ \pi^2}{2L^4}\Big[ \bar r L^2- \frac{(d+2)\pi^2}{4}\Big]\Bigg\}\nonumber\\&& +\;\;\frac{  A_{d-1}\; \bar r^{(d-3)/2}}{2(d-1)(5-d)}\;L^{-3}\Big[ \bar r L^2- \frac{(d-1)\pi^2}{2}\Big]  \nonumber\\
&&+ \;L^{-d}{\cal P}(\bar rL^2,\rho ) ,
\end{eqnarray}
\begin{subequations}
\label{bulksing}
\begin{align}
\label{a}
f^+_{b,s}(r_0)=-\frac{A_d}{d\varepsilon}\;r_0^{d/2},\hspace{4.0cm}
\\
\label{b}
f^-_{b,s}(r_0)=-\frac{r_0^2}{24u_0}-\frac{A_d}{d\varepsilon}\;(-r_0)^{d/2}\big[3-\frac{\varepsilon(d+2)}{4}\big]
\end{align}
\end{subequations}
for $r_0>0$ and $r_0<0$, respectively,
apart from terms of $O(M_0^2,\rho^{d-1})$, with $\bar r= \hat r+\pi^2/L^2$ and $A_d = \Gamma(3-d/2)[2^{d-2} \pi^{d/2}(d-2)]^{-1}$. The function ${\cal P}$ is given by
\begin{eqnarray}
\label{calP}
&&{\cal P}(\bar rL^2,\rho )= \frac{1}{2^{d+1}\pi} \int\limits_0^\infty
dz
   \Bigg\{
  \left(\frac{\pi}{z}\right)^{1/2}\left(1+z+\frac{z^2}{2}\right)\nonumber\\ &&
  - \Big[2\rho K(4\rho^2z)\Big]^{d-1}e^{z}[K(z)-1]\left(\frac{\pi}{z}\right)^{(1-d)/2}\nonumber\\ && - \left(1+z\right)\Bigg\}  \left(\frac{\pi}{z}\right)^{(d+1)/2} \exp {\left[-\bar rL^2z /\pi^2\right]}
\end{eqnarray}
with $ K(z) = \sum^{\infty}_{m= -\infty} \;\exp (- z m^2)$ and $2\rho K(4\rho^2z)\to (\pi/z)^{1/2}$ for $\rho \to 0$.
For finite $\bar rL^2>0$, the integral ${\cal P}(\bar rL^2,\rho)$ exists in $1<d<5$ dimensions for $\rho\geq0$. Note that
a $d=3$ pole is contained in the surface contribution $\propto A_{d-1}$ in Eq. (\ref{excess-free-energy}). This pole term is well understood as an artifact of perturbation theory due to the vanishing of the critical exponent of the Gaussian surface energy density at $d=3$ \cite{chen-dohm2003}. Here the $d=3$ pole is not problematic because it is canceled in the quantity $ - f^{{\text ex}} - L\partial f^{{\text ex}}/\partial L = F_{{\text Cas}}$. Our function $f_s$, Eq. (\ref{excess-free-energy}), is nonsingular at $r_0=0$ for finite $L$, in agreement with general analyticity requirements \cite{fussnote}. In fact, $f_s(r_0,L,L_\parallel)$ constitutes the analytic continuation of an earlier result (as given by the singular part of Eqs. (66),(67), and (69) of Ref. \cite{chen-dohm2003} that is valid only for $r_0\geq0$) to the region $r_0>-\pi^2/L^2$ \cite{analytic}.  A proof of this statement will be given elsewhere.

The conceptual progress of our approach manifests itself in the representation of Eqs. (\ref{excess-free-energy}) and (\ref{calP}) in terms of closed functions of the {\it shifted} variable ${\bar r}$ relative to the (unrenormalized) {\it film} critical point  rather than relative to bulk $T_c$. The variable $\bar r$ is  an analytic function of $r_0$ which immediately proves the analyticity of $f_s$ \cite{Krech-Dietrich}.
Equations (\ref{excess-free-energy})  - (\ref{calP}) do not yet correctly describe the finite-size scaling behavior in terms of the scaling variable $\tilde x$ with the correct critical exponent $\nu$. This will be achieved by appropriate renormalizations that we perform within the minimal subtraction scheme at fixed dimension $d$ \cite{dohm2008,dohm1985}.

It is straightforward to extend our calculation to $n > 1$ as far as the ${\it disordered}$ phase above $T_{c,\text{film}}$ is concerned. Then $n-1$ transverse contributions  exist which depend on the "transverse" parameter  $ \tilde r_{{\rm T}}(M_0^2) = r_0+4u_0 M_0^2$ rather than then "longitudinal" parameter $ \tilde r(M_0^2)$ defined above. These definitions are parallel to those in Ref. \cite{dohm2013}. Especially in the limit $\rho \to 0$, each of the $n$ components of $\varphi({\bf x})$ contributes equally to the free energy which amounts to multiplying both $f_s$ and the bulk part $f^+_{b,s}$ by $n$. As far as the transverse finite-size contributions are concerned, our approach is not applicable to $T< T_{c,\text{film}}(L)$ where $ \tilde r_{{\rm T}}(M_0^2) $ would become negative.
This can be traced back to the factor $3/2$ in the $\Phi^4$ term of $H_0$.
As far as the transverse bulk contribution is concerned we argue, however, that no unrenormalized transverse bulk contributions below $T_c$ exists at $O(u_0^{-1})$ and $O(1)$ as is known from bulk perturbation theory \cite{str}. This remains true also for the renormalized bulk theory in terms of the renormalized coupling $u$.
Thus, on our level of the theory which neglects terms of $O(u)$ and for the application restricted to $T_{c,\text{film}}(L)<T<T_c $, we approximate the bulk part below $T_c$ for general $n \geq 1$ by the  longitudinal bulk contribution below $T_c$ as given in Eq. (\ref{brenorm}) below (where the $n$ dependence enters only through the fixed point value $u^*$ and the flow parameter $l_-$).

The quantity of primary interest is the Casimir force scaling function $X(\tilde x)=\lim_{\rho \to 0} X(\tilde x,\rho)$ in the film limit. From Eqs. (\ref{excess-free-energy})  - (\ref{calP})  we derive its analytic form for general $n$  above $(+)$ and below $(-)$ $T_c$ and above $T_{c,\text{film}}(L)$
\begin{eqnarray}
\label{Casimir-film} &&X(\tilde x)=  -A_d l_\pm^d\;\frac{n}{\varepsilon}\Big[
\frac{1}{4}-\frac{l_\pm^\varepsilon}{d}\hat l_\pm^{-\varepsilon}\Big] +   F^\pm_b(\tilde x) \;\nonumber\\ && +\frac{A_d n\pi^2l_\pm^4}{d}\hat l_\pm^{d-6} + \; \frac{A_d n\pi^2}{2d}\hat l_\pm^{-\varepsilon}\Bigg\{ -\hat l_\pm^2 + \pi^2(14-d)/4  \nonumber\\
&&+ \frac{\pi^4}{4\hat l_\pm^2}(d-4)(d+2)\Bigg\}   -n\frac{\pi^{(9-d)/2}}{2^{d}}\Gamma\Big(\frac{5-d}{2}\Big)\hat l_\pm^{d-5}\nonumber\\
&& +\frac{n}{2^{d+1}\pi} \int\limits_0^\infty
dz \Big[d-1\pm 2\frac{l_\pm^2}{\pi^2}z\Big]
   \Bigg\{
  \left(\frac{\pi}{z}\right)^{1/2}\left(1+z+\frac{z^2}{2}\right)\nonumber\\ &&
  - e^{z}[K(z)-1]- \left(1+z\right)\Bigg\}  \left(\frac{\pi}{z}\right)^{(d+1)/2} \exp {\left[-\hat l_\pm^2z /\pi^2\right]},\nonumber\\
\end{eqnarray}
\begin{subequations}
\label{bulksingrenorm}
\begin{align}
\label{arenorm}
F^+_b(\tilde x) =-A_d l_+^d n/(4d),\hspace{2.8 cm}
\\
\label{brenorm}
F^-_b(\tilde x)=-A_d l_-^d[1/(24u^*)+1/(4d)-1/4],
\end{align}
\end{subequations}
where $l_\pm=(\pm\; \tilde x Q^*)^\nu$ and $\hat l_\pm =(\pm \;l_\pm^2 + \pi^2)^{1/2}$. The quantity $ Q^* =Q(1,u^*,d) $ is the fixed point value of the $n$ dependent amplitude function $Q (1, u, d)$ of the second-moment bulk correlation length above $T_c$ \cite{dohm1985}.
\begin{figure}[!ht]
\begin{center}
\subfigure{\includegraphics[clip,width=7.92cm]{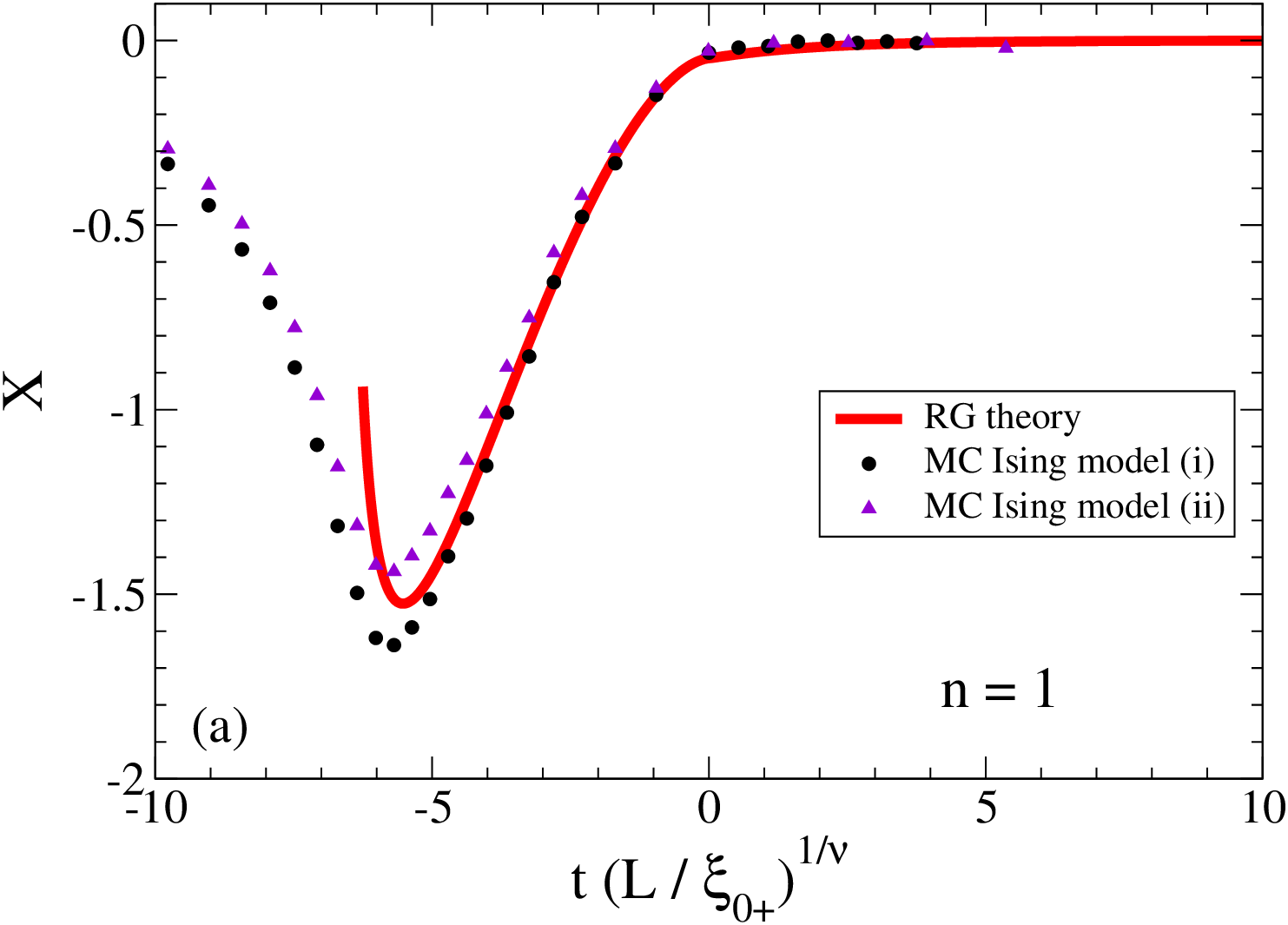}}
\subfigure{\includegraphics[clip,width=7.92cm]{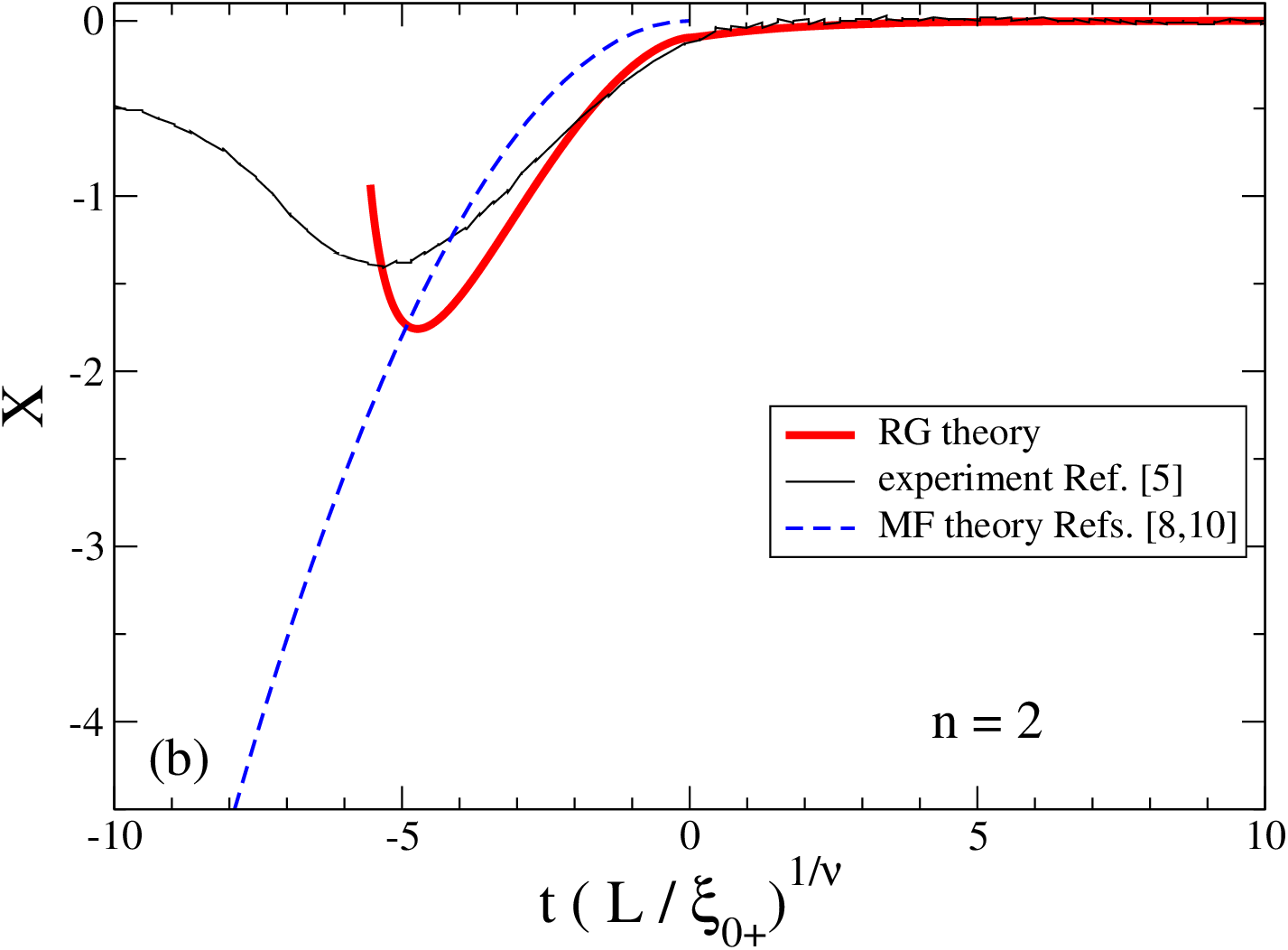}}
\subfigure{\includegraphics[clip,width=7.92cm]{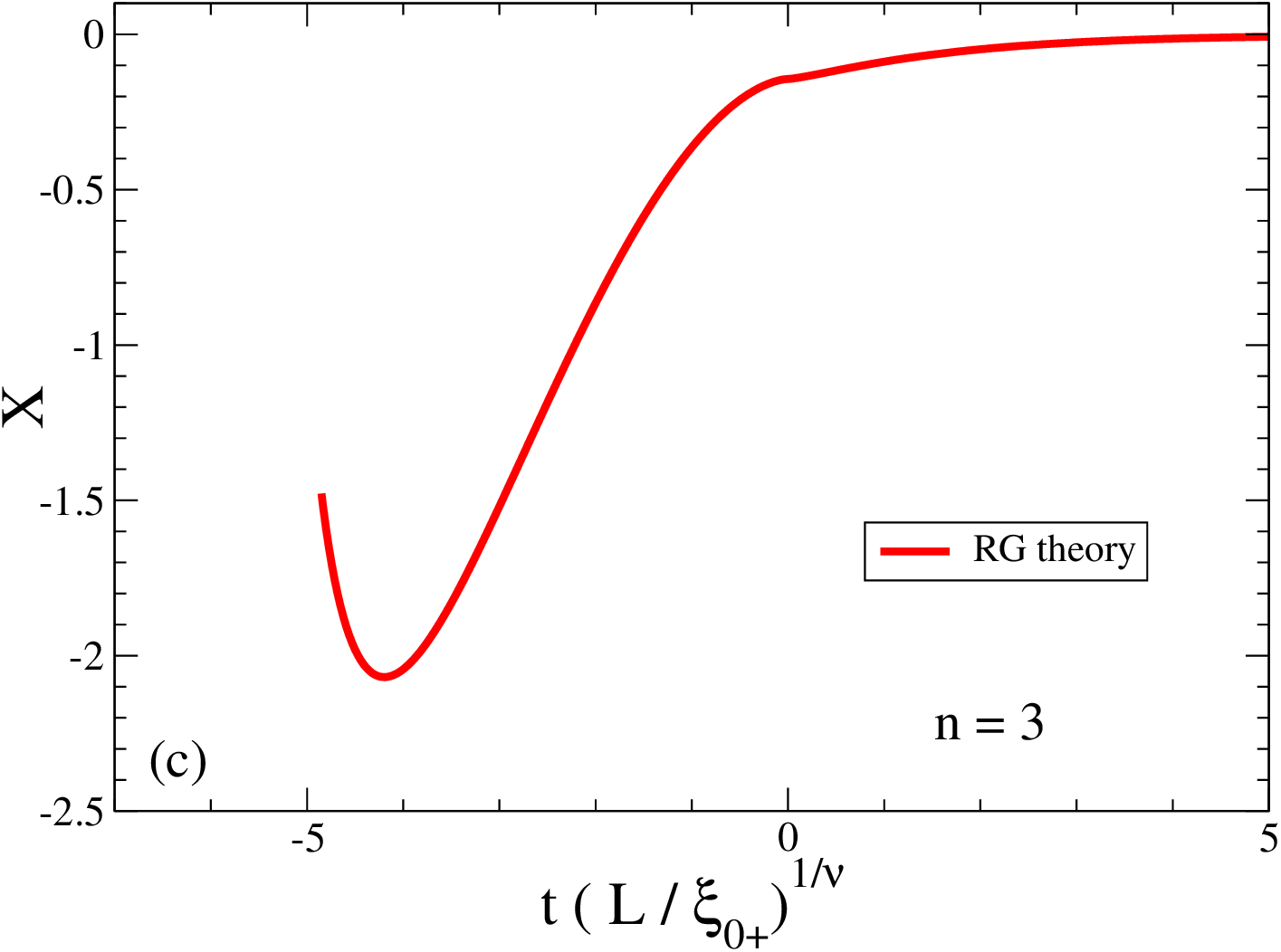}}
\end{center}
\caption{(Color online) Scaling function $X(\tilde x)$
of the Casimir force as a function of $\tilde x = t (L/\xi_{0+})^{1/\nu}$ in three dimensions
for $n=1,2,3$. Thick solid lines: RG theory  from Eqs. (\ref{Casimir-film}) and (\ref{bulksingrenorm}).  MC data (i) and (ii) in panel (a) from Ref. \cite{vasilyev2007} for  the Ising model with $L=20$. Thin line in panel (b): $^4$He data from Ref. \cite{garcia} with $\xi_{0+}=1.43$ $10^{-8}$ cm.  Dashed line in panel (b): RG improved MF theory from Refs. \cite{zandi2007,maciolek} with a minimum $ X^{{\text MF}}_{{\text min}} =-6.92$ at $\tilde x^{{\text MF}}_{{\text min}} =-9.87$.}
\end{figure}
Equations (\ref{Casimir-film}) and (\ref{bulksingrenorm}) are the central  result of this paper. They contain no adjustable parameters. They are valid in $2<d<4$ dimensions (with a finite limit for $d\to 4$) including $d=3$ in the range $\tilde x > \tilde x_{c,{\text film}}$ which is the renormalized counterpart of the range $r_0> -\pi^2/L^2 $ mentioned above.  The film transition occurs at  $ \hat l_-=0$ or  $ l_- =\pi$, i.e.,
\begin{eqnarray}
\label{film}
\tilde x_{c,{\text film}}=-\pi^{1/\nu}/Q^*
\end{eqnarray}
The crucial conceptual advance of our theory is the function $X$ below $T_c$ which provides a description relative to the renormalized film critical point (\ref{film}) as reflected in the variable $ \hat l_-=( -l_-^2 + \pi^2)^{1/2}$.  Our function $X(\tilde x)$ is an analytic function in the entire region $\tilde x_{c,{\text film}}<\tilde x <0$ and $0< \tilde x <\infty$. By definition, $X(\tilde{x})$ has a weak singularity at $\tilde x=0$ arising from the bulk part of $f^{{\text ex}}$.

Our result in Eqs. (\ref{Casimir-film}) and (\ref{bulksingrenorm}) is compared with experimental and MC data in Figs. 1 (a) and (b). For $d=3$ we employ the following numerical values \cite{dohm1985,larin,pelissetto,CDS1996}  $u^*=0.0404, 0.0362$, $Q^*=0.946, 0.939$, and $\nu= 0.6301, 0.671$ for $n=1,2$, respectively.
We obtain $\tilde x_{c,{\text film}}=-6.44, -5.86$ for $n=1,2$, respectively.  This is not far from the observed transitions at $\tilde x_{c,{\text film}}= -7.6$ for both the Ising $ (n = 1)$ and the $ XY (n =  2)$ universality classes \cite{vasilyev2007}.
The positions of the minima predicted by our theory are  $ \tilde x_{{\text min}} =-5.53,-4.73$ for $n=1,2$, respectively. This is in excellent agreement with the position $ \tilde x^{{\text MC}}_{{\text min}}=-5.7$ observed by MC simulations for $n=1$ \cite{vasilyev2007} and in reasonable agreement with $ \tilde x^{{\text exp}}_{{\text min}}=-5.7$  measured by experiments for $n=2$ \cite{garcia}, as shown in Figs. 1 (a) and (b). The position predicted by MF theory  \cite{zandi2007,maciolek} $\tilde x^{{\text MF}}_{{\text min}}= - \pi^2 =-9.87$ differs considerably from the observed position. Also the shape of $X(\tilde x)$ and the depth of the minimum $ X_{{\text min}} =-1.53$ predicted by our theory for $n=1$ are in excellent agreement with the MC data (Fig. 1 (a))  while semiquantitative agreement with the experimentally measured depth for $n = 2$ (Fig. 1 (b))  is found. The RG improved MF theory \cite{zandi2007} (dashed line in Fig. 1 (b) ) has a minimum $ X^{{\text MF}}_{{\text min}} =-6.92$ that is far from the experimental value and is outside the range of the vertical scale shown in Fig. 1 (b).

Although our theory captures well the film transition point Eq. (\ref{film}), our function $X$ does not correctly describe the weak singularity at $\tilde x_{c,{\text film}}$ \cite{vasilyev2007} for $n=1,2$ but yields a divergence  for $n\geq1$. Nevertheless we expect that our function $X(\tilde x)$ provides a reasonable prediction at a semiquantitative level for general $n>2$ in the range $\tilde x_{{\text min}}\lesssim \tilde x\leq \infty$. An application of our result in Eqs. (\ref{Casimir-film}) and (\ref{bulksingrenorm}) to $n=3$ (with parameters $u^*=0.0327, Q^*= 0.937, \nu=0.7112 $ taken from Refs. \cite{campostrini,CDS1996,larin}) yields a pronounced minimum $X_{{\text min}}=-2.07$ at $\tilde x_{{\text min}}= -4.20$ as shown in Fig. 1 (c). The latter value is close to  $\tilde x^{(\infty)}_{{\text min}}= -4.56$ of the pronounced minimum found recently in the large - $n$ limit \cite{diehl2012}. It would be interesting to test our $n=3$ prediction by MC simulations for Heisenberg models with free BC.

Our analytic theory provides the opportunity of studying separately the contributions arising from bulk and finite-size parts. For  $n=1,2,3$, the occurrence of the pronounced minimum can be understood as the result of a competition between a decreasing bulk contribution [as represented by the term  $F^-_b(\tilde x)$ in Eq. (\ref{Casimir-film})] and an increasing $L$ - dependent fluctuation contribution to the Casimir force as the temperature is lowered below bulk $T_c$ toward $T_{c,\text{film}}(L)$. An analysis of Eqs. (\ref{Casimir-film}) and (\ref{bulksingrenorm}) for larger $n>3$ can answer the question whether this feature persists up to $n=\infty$ \cite{diehl2012}. The fluctuation contribution is missing in MF theory which explains why no minimum exists in MF theory (dashed line in Fig. 1 (b)) above the MF film transition temperature $ \tilde x^{{\text MF}}_{c,{\text film}}= - \pi^2 =-9.87$.

To summarize, we have shown that the pronounced minima of the Casimir force scaling function of $O(n)$ symmetric film systems observed in experiments \cite{garcia} and MC simulations \cite{hucht,vasilyev2007,hasenbusch2010} can be described analytically within a finite-size RG approach on the basis of the $\varphi^4$ model with Dirichlet BC. Our approach may  also be applicable to the low-temperature phase of superfluid films and superconducting films where Goldstone modes play an important role \cite{zandi2004,dohm2013}.

The author is grateful to O. Vasilyev for providing  the data of Refs. \cite{vasilyev2007,garcia} in numerical form.


\begin{thebibliography}{99}
\bibitem{analytic}
For $\rho=0$, a different representation of $f_s$ in the range $ r_0 \geq -\pi^2/L^2$ was given in Eqs. (4.16) and (4.17b) of Ref. \cite{kastening-dohm} where the integral (4.17b) is divergent for both $d=3$ and $d=4$, in contrast to our integral $\cal P$.
\bibitem{Krech-Dietrich}
For $\rho=0$ and $n=1$, the scaling function of the finite-size part of $f_s$ below $T_c$ has been predicted in Fig. 4 of Ref. \cite{KrDi92} (a). It is applicable only very close to $T_c$. Analyticity of the free energy at $T_c$ has been invoked as a requirement but no proof of analyticity has been given in Ref. \cite{KrDi92} (a) due to the truncation of the series in Eq. (7.1).
\end{thebibliography}
\end{document}